\documentclass[12pt]{article}

\usepackage{amsmath}
\usepackage{amssymb}
\usepackage{cite}
\usepackage{hyperref}

\begin{document}

\title{Gauge-invariant description of several $(2{+}1)$-dimensional integrable nonlinear evolution equations}

\author{V.~G. Dubrovsky and A.~V. Gramolin
\\{\small Novosibirsk State Technical University, Novosibirsk, Russia}
\\{\small E-mail: \href{mailto:dubrovsky@academ.org}{dubrovsky@academ.org} and \href{mailto:gramolin@gmail.com}{gramolin@gmail.com}}}

\date{}

\maketitle

\begin{abstract}
We obtain new gauge-invariant forms of two-dimensional integrable systems of nonlinear equations: the Sawada--Kotera and Kaup--Kuperschmidt system, the generalized system of dispersive long waves, and the Nizhnik--Veselov--Novikov system. We show how these forms imply both new and well-known two-dimensional integrable nonlinear equations: the Sawada--Kotera equation, Kaup--Kuperschmidt equation, dispersive long-wave system, Nizhnik--Veselov--Novikov equation, and modified Nizhnik--Veselov--Novikov equation. We consider Miura-type transformations between nonlinear equations in different gauges.
\end{abstract}
\vspace{7mm}

\section{Introduction}

The fundamental methods based on gauge transformations and the concept of gauge invariance are currently widely used in physics and mathematics, in particular, in the theory of integrable nonlinear equations. The first applications of these methods in the theory of integrable nonlinear equations were proposed in~\cite{Zakharov&Shabat_1974, Kuznetsov&Mikhailov_1977, Zakharov&Mikhailov_1978, Zakharov&Takhtadzhyan_1979, Konopelchenko_PLA_1982, Konopelchenko&Dubrovsky_PLA_1983, Konopelchenko&Dubrovsky_1984} (also see~\cite{Theory_of_solitons, Faddeev&Takhtadzhyan, Ablowitz&Clarkson, Konopelchenko_book_1, Konopelchenko_book_2, Konopelchenko_book_3} and the references therein).

A great many gauge-equivalent pairs of integrable models have been found. In the one-dimensional case, these include the nonlinear Schr\"{o}dinger equation and Heisenberg ferromagnet equation, the equations describing the massive Thirring model and the two-dimensional relativistic field theory, and the Korteweg--de~Vries (KdV) and modified KdV (mKdV) equations. In the two-dimensional case, the currently most widely known gauge-equivalent pairs of equations are the Kadomtsev--Petviashvili (KP) and modified KP (mKP) equations and the Davey--Stewartson and Ishimori equations (see~\cite{Theory_of_solitons, Faddeev&Takhtadzhyan, Ablowitz&Clarkson, Konopelchenko_book_1, Konopelchenko_book_2, Konopelchenko_book_3, Konopelchenko&Rogers_1992} and the references therein).

We introduce the required terms and illustrate the unifying role of gauge transformations and gauge invariance with the well-known example of the interaction between a charged spinless particle and an external electromagnetic field with the vector and scalar potentials $\vec A(\vec r, t)$ and $\psi(\vec r, t)$,
\begin{equation}
i \hbar \psi_t = \frac{\bigl(\, \widehat {\vec p} - q \vec A \, \bigr)^2}{2 m} \psi + q \phi \psi, \label{eq.1.1}
\end{equation}
where $\widehat{\vec{p}} = -i \hbar \vec{\nabla}$ is the particle momentum operator. The particle coupling to external fields has the well-known gauge-invariant form.

From the standpoint of the inverse scattering method, Eq.~(\ref{eq.1.1}) is an auxiliary linear problem: a linear partial differential equation with variable coefficients for the wave function~$\psi$. Under the gauge transformation
\begin{equation}
\psi \rightarrow \psi' = g^{-1} \psi, \qquad \psi = g \psi' = \exp\left[\frac{i\chi(\vec r, t) q}{\hbar}\right] \psi', \label{eq.1.2}
\end{equation}
Eq.~(\ref{eq.1.1}) preserves its form if the potentials $\vec A$ and $\phi$ are transformed as
\begin{equation}
\vec A \rightarrow \vec A' = \vec A - \vec \nabla \chi, \qquad \phi \rightarrow \phi' = \phi + \chi_t. \label{eq.1.3}
\end{equation}

Eliminating the gauge function $\chi$ from~(\ref{eq.1.3}), we obtain the relations
\begin{equation}
[\vec \nabla \times \vec A'] = [\vec \nabla \times \vec A], \qquad -\vec \nabla \phi' - \frac{\partial \vec A'}{\partial t} = -\vec \nabla \phi - \frac{\partial \vec A}{\partial t}, \label{eq.1.4}
\end{equation}
which mean that the quantities
\begin{equation}
\vec B \stackrel{\mathrm{def}}{=} [\vec \nabla \times \vec A], \qquad \vec E \stackrel{\mathrm{def}}{=} -\vec \nabla \phi - \frac{\partial \vec A}{\partial t}, \label{eq.1.5}
\end{equation}
which are well known in electrodynamics as the magnetic field induction $\vec B$ and electric field strength $\vec E$, are invariants of gauge transformations. Moreover, definitions~(\ref{eq.1.5}) for the invariants $\vec B$ and $\vec E$ imply the subsystem of equations
\begin{equation}
\text{div} \vec B = \vec \nabla \cdot \vec B = 0, \qquad \vec \nabla \times \vec E = -\frac{\partial \vec B}{\partial t}, \label{eq.1.6}
\end{equation}
which is a (sourceless) subsystem of the fundamental system of Maxwell equations. Thus, starting from the principle of local gauge invariance, we can obtain the gauge-invariant nonstationary Schr\"{o}dinger equation for a charged spinless particle in an external electromagnetic field, the gauge invariants $\vec B$ and $\vec E$, and a (sourceless) subsystem of fundamental Maxwell equations. Similar considerations are applicable to the case of integrable nonlinear equations.

\section{Gauge-invariant integrable system of KP and mKP equations}
\setcounter{equation}{0}

We illustrate the methods based on the notion of gauge invariance with the well-known example of an integrable system of nonlinear KP and mKP equations. Auxiliary linear problems for the KP and mKP equations are particular cases of the following linear problems (with the respective gauges $C \left(u_0 \neq 0, u_1 = 0\right)$ and $C \left(u_0 = 0, u_1 \neq 0\right)$ for the KP and mKP equations):
\begin{align}
&L_1 \psi = \bigl(\sigma \partial_y + \partial_x^2 + u_1 \partial_x + u_0\bigr) \psi = 0, \label{eq.2.1} \\
&L_2 \psi = \bigl(\partial_t + 4\partial_x^3 + v_2 \partial_x^2 + v_1 \partial_x + v_0\bigr) \psi = 0, \label{eq.2.2}
\end{align}
where the constant values $\sigma = i$ or $\sigma^{2} = -1$ and $\sigma = 1$ or $\sigma^{2} = 1$ respectively correspond to the cases of KP--I or mKP--I and KP--II or mKP--II equations. Because of~(\ref{eq.2.1}) and (\ref{eq.2.2}), the compatibility condition in the Lax form $[L_1, L_2] = 0$ leads to a system of evolution equations for the field variables $u_1$ and $u_0$~\cite{Konopelchenko_PLA_1982, Konopelchenko&Dubrovsky_1984},
\begin{equation}
\begin{split}
&u_{1t} + u_{1xxx} - \frac{3}{2} u_1^2 u_{1x} - 3\sigma u_{1x} \partial_x^{-1} u_{1y} + 3\sigma^2 \partial_x^{-1} u_{1yy} + {} \\
&\qquad {} + 6u_{0xx} - 6\sigma u_{0y} + 6 \bigl(u_0 u_1\bigr)_x - 2v_{0x} = 0, \\
&u_{0t} + 4\sigma u_{0xxx} + 6u_0 u_{0x} + 6u_1 u_{0xx} + \frac{3}{2} u_1^2 u_{0x} + 3u_{0x} u_{1x} - {} \\
&\qquad {} - 3\sigma u_{0x} \partial_x^{-1} u_{1y} - u_1 v_{0x} - v_{0xx} - \sigma v_{0y} = 0.
\end{split} \label{eq.2.3}
\end{equation}
For the gauge $C \left(u_0 \neq 0, u_1 = 0\right)$ with the choice of the variable $v_0 = 3u_{0x} - 3\sigma \partial_x^{-1} u_{0y}$, system~(\ref{eq.2.3}) implies the KP equation~\cite{Zakharov&Shabat_1974, Dryuma_1974}
\begin{equation}
u_{0t} + u_{0xxx} + 6u_0 u_{0x} + 3\sigma^2 \partial_x^{-1} u_{0yy} = 0. \label{eq.2.4}
\end{equation}
For the gauge $C \left(u_0 = 0, u_1 \neq 0\right)$ with the choice of the variable $v_0 = \text{const}$, system~(\ref{eq.2.3}) implies the mKP equation
\begin{equation}
u_{1t} + u_{1xxx} - \frac{3}{2} u_1^2 u_{1x} - 3\sigma u_{1x} \partial_x^{-1} u_{1y} + 3\sigma^2 \partial_x^{-1} u_{1yy} = 0. \label{eq.2.5} \\
\end{equation}
System of equations~(\ref{eq.2.3}) is called the system of KP--mKP equations.

In terms of the pure gauge variable $\rho$ and the invariant $w_0$ of the gauge transformations $\psi \rightarrow \psi' = g^{-1} \psi$ determined by the expressions
\begin{equation}
u_1(x, y, t) = 2\frac{\rho_x}{\rho}, \qquad w_0 = u_0 - \frac{1}{2} u_{1x} - \frac{1}{4} u_1^2 - \frac{\sigma}{2} \partial_x^{-1} u_{1y}, \label{eq.2.6}
\end{equation}
system of equations~(\ref{eq.2.3}) becomes
\begin{align}
&\rho_t + 4\rho_{xxx} + 6\rho_x w_0 + 3\rho w_{0x} - 3\sigma \partial_x^{-1} w_{0y} - \rho v_0 = 0, \label{eq.2.7} \\
&w_{0t} + w_{0xxx} + 6w_0 w_{0x} + 3 \sigma^2 \partial_x^{-1} w_{0yy} = 0. \label{eq.2.8}
\end{align}

System~(\ref{eq.2.7}), (\ref{eq.2.8}) is the manifestly gauge-invariant form of system of KP--mKP equations~(\ref{eq.2.3}). This system has the following structure. It contains gauge-invariant KP equation~(\ref{eq.2.8}) for the gauge invariant $w_0$ and Eq.~(\ref{eq.2.7}) for the pure gauge variable $\rho$ with additional terms containing the gauge invariant $w_0$ and the additional field variable $v_0$. If the invariant is zero, $w_0 = 0$, then KP--mKP system~(\ref{eq.2.7}), (\ref{eq.2.8}) reduces to the linear evolution equation for the pure gauge variable $\rho$:
\begin{equation}
\rho_t + 4\rho_{xxx} - \rho v_0 = 0. \label{eq.2.9}
\end{equation}

We use the gauge invariant $w_0$ to obtain the Miura-type transformation from Eq.~(\ref{eq.2.8}),
\begin{equation}
w_0 = u_0 = -\frac{1}{2} u_{1x} - \frac{1}{4} u_1^2 - \frac{\sigma}{2} \partial_x^{-1} u_{1y}, \label{eq.2.10}
\end{equation}
which relates the solutions $u_0$ (for the gauge $C \left(u_0 \neq 0, u_1 = 0\right)$) and $u_1$ (for the gauge $C \left(u_0 = 0, u_1 \neq 0\right)$) of KP equations~(\ref{eq.2.4}) and mKP equations~(\ref{eq.2.5}). This also follows from Eq.~(\ref{eq.2.8}) written for the gauges  $C \left(u_0 \neq 0, u_1 = 0\right)$ and $C \left(u_0 = 0, u_1 \neq 0\right)$:
\begin{multline}
u_{0t} + u_{0xxx} + 6u_0 u_{0x} + 3\sigma^2 \partial_x^{-1} u_{0yy} = {} \\
{} - \frac{1}{2} \Bigl(\partial_x + \frac{1}{2} u_1 + \sigma \partial_x^{-1} \partial_y\Bigr) \Bigl(u_{1t} + u_{1xxx} - \frac{3}{2} u_1^2 u_{1x} - 3\sigma u_{1x} \partial_x^{-1} u_{1y} + 3\sigma^2 \partial_x^{-1} u_{1yy}\Bigr) = 0. \label{eq.2.11}
\end{multline}

We stress that the separation of the physical and the pure gauge degrees of freedom in integrable nonlinear equations and their gauge-invariant formulation can be used to study the structure of these equations and the relations between their different gauge-equivalent realizations.

\section{Manifestly gauge-invariant integrable system of two-dimensional Kaup--Kuperschmidt and Sawada--Kotera equations}
\setcounter{equation}{0}

In this section, we briefly discuss the results obtained up to now concerning the manifestly gauge-invariant formulation of two-dimensional integrable generalizations of the Kaup--Kuperschmidt ($2D$KK) and Sawada--Kotera ($2D$SK) nonlinear equations. The auxiliary linear problems for these equations are particular cases (in the gauge $C \left(u_2 = 0, u_1 \neq 0, u_0 = u_{1x}/2\right)$ for the $2D$KK equation and in the gauge $C \left(u_2 = 0, u_1 \neq 0, u_0 = 0\right)$ for the $2D$SK equation) of the problems~\cite{Konopelchenko&Dubrovsky_1984}
\begin{equation}
\begin{split}
&L_1 \psi = \bigl(\sigma \partial_y + \partial_x^3 + u_2 \partial_x^2 + u_1 \partial_x + u_0\bigr) \psi = 0, \\
&L_2 \psi = \bigl(\partial_t +\kappa \partial_x^5 + v_4 \partial_x^4 + v_3 \partial_x^3 + v_2 \partial_x^2 + v_1 \partial_x + v_0\bigr) \psi = 0.
\end{split} \label{eq.3.1}
\end{equation}

In terms of the pure gauge variable $\rho$ and the invariants $w_0$ and $w_1$ of the gauge transformations $\psi\rightarrow\psi'=g^{-1}\psi$ given by the expressions
\begin{align}
&u_2 = 3\frac{\rho_x}{\rho}, \qquad w_1 = u_1 - u_{2x} - \frac{1}{3} u_2^2, \label{eq.3.2} \\
&w_0 = u_0 - \frac{1}{3} u_1 u_2 - \frac{1}{3} u_{2xx} + \frac{2}{27} u_2^3 - \frac{\sigma}{3} \partial_x^{-1} u_{2y}, \label{eq.3.3}
\end{align}
we obtain the manifestly gauge-invariant integrable system of $2D$KK--$2D$SK nonlinear equations
\begin{align}
&\rho_{t} + \kappa \rho_{xxxxx} - \rho v_0 + \frac{5}{3} \kappa \bigl(\rho_{xx} w_1\bigr)_x + \frac{5}{3} \kappa \bigl(\rho_x w_0\bigr)_x + \frac{5}{9} \kappa \rho_x w_1^2 + \frac{10}{9} \kappa \rho_x w_{1xx} + {} \nonumber \\
&\qquad {} + \frac{10}{9} \kappa \rho w_{0xx} + \frac{10}{9} \kappa \rho w_0 w_1 - \frac{5}{9} \kappa \sigma \rho_x \partial_x^{-1} w_{1y} - \frac{5}{9} \kappa \sigma \rho \partial_x^{-1} w_{0y} = 0, \label{eq.3.4} \\
&w_{1t} - \frac{1}{9} \kappa w_{1xxxxx} - \frac{5}{9} \kappa \bigl(w_1 w_{1xx}\bigr)_x - \frac{5}{3} \kappa \bigl(w_0 w_{1x}\bigr)_x - \frac{5}{9} \kappa w_1^2 w_{1x} + \frac{10}{3} \kappa w_0 w_{0x} - {} \nonumber \\
&\qquad {} - \frac{5}{9} \kappa \sigma w_{1xxy} - \frac{5}{9} \kappa \sigma w_1 w_{1y} + \frac{5}{9} \kappa \sigma^2 \partial_x^{-1} w_{1yy} - \frac{5}{9} \kappa \sigma w_{1x} \partial_{x}^{-1} w_{1y}=0, \label{eq.3.5}
\\
&w_{0t} - \frac{1}{9} \kappa w_{0xxxxx} - \frac{5}{9} \kappa \bigl(w_0 w_1\bigr)_{xxx} - \frac{5}{9} \kappa \bigl(w_0 w_{1xx}\bigr)_x + \frac{5}{3} \kappa \bigl(w_0 w_{0x}\bigr)_x - \frac{5}{9} \kappa \bigl(w_0 w_1^2\bigr)_x - {} \nonumber \\
&\qquad {} - \frac{5}{9} \kappa \sigma w_{0xxy} - \frac{10}{9} \kappa \sigma w_0 w_{1y} - \frac{5}{9} \kappa \sigma w_1 w_{0y} + \frac{5}{9} \kappa \sigma^2 \partial_x^{-1} w_{0yy} - \frac{5}{9} \kappa \sigma w_{0x} \partial_x^{-1} w_{1y}=0. \label{eq.3.6}
\end{align}
This system consists of gauge-invariant system of equations~(\ref{eq.3.5}), (\ref{eq.3.6}) for the gauge invariants~$w_0$ and $w_1$ and Eq.~(\ref{eq.3.4}) for the pure gauge variable~$\rho$ with additional terms containing the gauge invariants~$w_0$ and $w_1$ and the additional variable $v_0$. For the zero-valued invariants $w_0 = 0$ and $w_1 = 0$, the $2D$KK--$2D$SK system given by~(\ref{eq.3.5}) and (\ref{eq.3.6}) reduces to the linear evolution equation for the pure gauge variable~$\rho$:
\begin{equation}
\rho_t + \kappa \rho_{xxxxx} - \rho v_0 = 0. \label{eq.3.7}
\end{equation}
Gauge-invariant system of equations~(\ref{eq.3.5}), (\ref{eq.3.6}) for the invariants $w_0$ and $w_1$ coincides in form with the system obtained in~\cite{Konopelchenko&Dubrovsky_1984}. This system admits the following reductions:
\begin{enumerate}
\item In the case $w_0 = 0$, we obtain an equation for the invariant~$w_1$ exactly coinciding with the $2D$SK equation~\cite{Konopelchenko&Dubrovsky_PLA_1984, Konopelchenko&Dubrovsky_1984}.
\item In the case $w_0 = w_{1x}/2$, we obtain an equation for the invariant~$w_1$ exactly coinciding with the $2D$KK equation~\cite{Konopelchenko&Dubrovsky_PLA_1984, Konopelchenko&Dubrovsky_1984}.
\end{enumerate}
Obviously, the $2D$KK and $2D$SK equations, as equations belonging to different sets of invariants, are not gauge-equivalent to each other.

\section{A manifestly gauge-invariant integrable two-dimensional generalized dispersive long-wave system}
\setcounter{equation}{0}

The gauge-invariant formulation of integrable systems of nonlinear equations can be obtained in all cases where the gauge freedom is taken into account in the corresponding auxiliary linear problems. In~\cite{Dubrovsky&Gramolin}, such a formulation was obtained for the auxiliary linear problems
\begin{align}
&L_1 \psi = \bigl(\partial_{\xi \eta}^2 + u_1 \partial_{\xi} + v_1 \partial_{\eta} + u_0\bigr) \psi = 0, \label{eq.4.1} \\
&L_2 \psi = \bigl(\partial_t + \kappa_1\partial_{\xi}^2 + \kappa_2\partial_{\eta}^2 + \tilde u_1 \partial_{\xi} + \tilde v_1 \partial_{\eta} + v_0\bigr) \psi = 0, \label{eq.4.2}
\end{align}
where $\kappa_1$ and $\kappa_2$ are constants, $\xi = x + \sigma y$ and $\eta = x - \sigma y$ are spatial variables, $\sigma^2 = \pm 1$, and the derivatives are $\partial_{\xi} = \partial / \partial \xi$, $\partial_{\eta} = \partial / \partial \eta$, $\partial_{\xi}^2 = \partial^2 / \partial \xi^2$, etc. Such a choice of auxiliary linear problems~(\ref{eq.4.1}) and (\ref{eq.4.2}) leads to the well-known two-dimensional generalized dispersive long-wave ($2D$gDLW) equation~\cite{Boiti&Leon&Pempinelli_1987}, the system of Davey--Stewartson (DS) equations~\cite{Davey&Stewartson_1974}, their reductions, and some other equations. All the listed well-known integrable nonlinear equations were previously obtained from the compatibility condition for auxiliary linear problems~(\ref{eq.4.1}) and (\ref{eq.4.2}) in the form of the Manakov triad representation~\cite{Manakov_1976}
\begin{equation}
[L_1, L_2] = B L_1. \label{eq.4.3}
\end{equation}

To obtain the manifestly gauge-invariant formulation of the corresponding integrable system of nonlinear equations, it is convenient to use the classical gauge invariants~$w_2$, $\widetilde w_2$, and $w_1$,
\begin{equation}
\begin{split}
&w_2 \stackrel{\mathrm{def}}{=} u_0 - u_{1\xi} - u_1 v_1 = u_0' - u_{1\xi}' - u_1' v_1', \\
&\widetilde w_2 \stackrel{\mathrm{def}}{=} u_0 - v_{1\eta} - u_1 v_1 = u_0' - v_{1\eta}' - u_1' v_1', \\
&w_1 \stackrel{\mathrm{def}}{=} u_{1\xi} - v_{1\eta} = u_{1\xi}' - v_{1\eta}',
\end{split} \label{eq.4.4}
\end{equation}
and the pure gauge variable $\rho$ related to the field variable $u_1(\xi,\eta,t)$ as
\begin{equation}
u_1 \stackrel{\mathrm{def}}{=} \left(\log{\rho}\right)_{\eta}. \label{eq.4.5}
\end{equation}
The variable $\rho$ corresponds to the pure gauge degrees of freedom and is transformed according to the simple law $\rho \rightarrow \rho' = g \rho$ under the gauge transformations $\psi\rightarrow\psi'=g^{-1}\psi$.

We note that the invariants $w_2$ and $\widetilde w_2$ in auxiliary problem~(\ref{eq.4.1}) are the Laplace invariants $h = w_2$ and $k = \widetilde w_2$ for the corresponding classical differential equation (see, e.g., the well-known Forsyth monograph on differential equations~\cite{Forsyth_book}).

In the case under study with auxiliary linear problem~(\ref{eq.4.2}) determining the time evolution, the corresponding integrable system of nonlinear equations in terms of the variables $\rho$, $w_1$, and $w_2$ becomes
\begin{align}
&\rho_t = -\kappa_1 \rho_{\xi \xi} - \kappa_2 \rho_{\eta \eta} - 2\kappa_1 \rho \partial_{\eta}^{-1} w_{2\xi} + 2\kappa_2 \rho_{\eta} \partial_{\xi}^{-1} w_1 + v_0 \rho, \label{eq.4.6} \\
&w_{1t} = -\kappa_1 w_{1\xi \xi} + \kappa_2 w_{1\eta \eta} - 2\kappa_1 w_{2\xi \xi} + 2\kappa_2 w_{2\eta \eta} - 2\kappa_1 \bigl(w_1 \partial_{\eta}^{-1} w_1\bigr)_{\xi} + 2\kappa_2 \bigl(w_1 \partial_{\xi}^{-1} w_1\bigr)_{\eta}, \label{eq.4.7} \\
&w_{2t} = \kappa_1 w_{2\xi \xi} - \kappa_2 w_{2\eta \eta} - 2\kappa_1 \bigl(w_2 \partial_{\eta}^{-1} w_1\bigr)_{\xi} + 2\kappa_2 \bigl(w_2 \partial_{\xi}^{-1} w_1\bigr)_{\eta}. \label{eq.4.8}
\end{align}
Gauge-invariant subsystem~(\ref{eq.4.7}), (\ref{eq.4.8}) for the invariants $w_1 = u_{1\xi} - v_{1\eta}$ and $w_2 = u_0 - u_{1\xi} - u_1 v_1$ of system of equations~(\ref{eq.4.6})--(\ref{eq.4.8}) with $u_1 = 0$, $v_1 = -q/2$, and $u_0 = (1 + r - q_{\eta})/4$ in terms of the variables $q$ and $r$ becomes
\begin{equation}
\begin{split}
&q_t = -\kappa_1 \partial_{\eta}^{-1} r_{\xi \xi} + \kappa_2 r_{\eta} - \frac{\kappa_1}{2} \bigl(q^2\bigr)_{\xi} + \kappa_2 q_{\eta} \partial_{\xi}^{-1} q_{\eta}, \\
&r_t = -\kappa_1 q_{\xi} + \kappa_2 \partial_{\xi}^{-1} q_{\eta \eta} -\kappa_1 q_{\eta \xi \xi} + \kappa_2 q_{\eta \eta \eta} -\kappa_1 \bigl(r q\bigr)_{\xi} + \kappa_2 \bigl(r \partial_{\xi}^{-1} q_{\eta}\bigr)_{\eta}.
\end{split} \label{eq.4.9}
\end{equation}
In the particular case where $\kappa_2 = 0$, system of equations~(\ref{eq.4.9}) reduces to the well-known integrable $2D$gDLW system~\cite{Boiti&Leon&Pempinelli_1987}
\begin{equation}
\begin{split}
&q_{t \eta} = -\kappa_1 r_{\xi \xi} - \frac{\kappa_1}{2} \bigl(q^2\bigr)_{\xi \eta}, \\
&r_{t \xi} = -\kappa_1 \bigl(q r + q + q_{\xi \eta}\bigr)_{\xi \xi}.
\end{split} \label{eq.4.10}
\end{equation}
In the one-dimensional limit $\xi = \eta$, system~(\ref{eq.4.9}) ($\kappa_1 - \kappa_2 = 1$) and system~(\ref{eq.4.10}) ($\kappa_1 = 1$) reduce to the well-known dispersive long-wave equation (see, e.g.,~\cite{Broer_1975}). Therefore, integrable system of nonlinear equations (\ref{eq.4.6})--(\ref{eq.4.8}) can be called the $2D$gDLW system.

Integrable system~(\ref{eq.4.6})--(\ref{eq.4.8}) in terms of the variables $\phi = \log{\rho}$, $w_1$, and $w_2$ becomes
\begin{align}
&\phi_t = -\kappa_1 \phi_{\xi \xi} - \kappa_2 \phi_{\eta \eta} - \kappa_1 (\phi_{\xi})^2 - \kappa_2 (\phi_{\eta})^2 - 2\kappa_1 \partial_{\eta}^{-1} w_{2\xi} + 2\kappa_2 \phi_{\eta} \partial_{\xi}^{-1} w_1 + v_0, \label{eq.4.11} \\
&w_{1t} = -\kappa_1 w_{1\xi \xi} - \kappa_2 w_{1\eta \eta} - 2\kappa_1 w_{2\xi \xi} + 2\kappa_2 w_{2\eta \eta} - 2\kappa_1 \bigl(w_1 \partial_{\eta}^{-1} w_1\bigr)_{\xi} + 2\kappa_2 \bigl(w_1 \partial_{\xi}^{-1} w_1\bigr)_{\eta}, \label{eq.4.12} \\
&w_{2t} = \kappa_1 w_{2\xi \xi} - \kappa_2 w_{2\eta \eta} - 2\kappa_1 \bigl(w_2 \partial_{\eta}^{-1} w_1\bigr)_{\xi} + 2\kappa_2 \bigl(w_2 \partial_{\xi}^{-1} w_1\bigr)_{\eta}. \label{eq.4.13}
\end{align}

Integrable system~(\ref{eq.4.6})--(\ref{eq.4.8}) in terms of the variables $\phi = \log{\rho}$, $w_2$, and $\widetilde w_2 = w_2 + w_1$ becomes more symmetric,
\begin{align}
&\phi_t = -\kappa_1 \phi_{\xi \xi} - \kappa_2 \phi_{\eta \eta} - \kappa_1 (\phi_{\xi})^2 - \kappa_2 (\phi_{\eta})^2 - 2\kappa_1 \partial_{\eta}^{-1} w_{2\xi} + 2\kappa_2 \phi_{\eta} \partial_{\xi}^{-1} w_1 + v_0, \label{eq.4.14} \\
&w_{2t} = \kappa_1 w_{2\xi \xi} - \kappa_2 w_{2\eta \eta} - 2\kappa_1 \bigl(w_2 \partial_{\eta}^{-1} (\widetilde w_2 - w_2)\bigr)_{\xi} + 2\kappa_2 \bigl(w_2 \partial_{\xi}^{-1} (\widetilde w_2 - w_2)\bigr)_{\eta}, \label{eq.4.15} \\
&\widetilde w_{2t} = - \kappa_1 \widetilde w_{2\xi \xi} + \kappa_2 \widetilde w_{2\eta \eta} - 2\kappa_1 \bigl(\widetilde w_2 \partial_{\eta}^{-1} (\widetilde w_2 - w_2)\bigr)_{\xi} + 2\kappa_2 \bigl(\widetilde w_2 \partial_{\xi}^{-1} (\widetilde w_2 - w_2)\bigr)_{\eta}. \label{eq.4.16}
\end{align}

All the mutually equivalent integrable systems of $2D$gDLW nonlinear equations considered above, (\ref{eq.4.6})--(\ref{eq.4.8}), (\ref{eq.4.11})--(\ref{eq.4.13}), and (\ref{eq.4.14})--(\ref{eq.4.16}) have the following common characteristic gauge structure:
\begin{itemize}
\item[\emph{a}.] They contain manifestly gauge-invariant subsystems~(\ref{eq.4.7}), (\ref{eq.4.8}) and (\ref{eq.4.12}), (\ref{eq.4.13}) of nonlinear equations for the gauge invariants $w_1$ and $w_2$ (or, equivalently, subsystem~(\ref{eq.4.15}), (\ref{eq.4.16}) for the gauge invariants $w_2$ and $\widetilde w_2$).
\item[\emph{b}.] They contain Eq.~(\ref{eq.4.6}) for the pure gauge variable $\rho$ (or Eq.~(\ref{eq.4.11}) for the variable $\phi = \log{\rho}$), which satisfies a simple transformation law $\rho \rightarrow \rho' = g \rho$ with additional terms containing the gauge invariants and the additional field variable $v_0$.
\end{itemize}
Such a structure of the $2D$gDLW system reflects a significant gauge freedom in auxiliary linear problems~(\ref{eq.4.1}) and (\ref{eq.4.2}).

We consider several particular gauges of systems of $2D$gDLW equations (\ref{eq.4.6})--(\ref{eq.4.8}), (\ref{eq.4.11})--(\ref{eq.4.13}), and (\ref{eq.4.14})--(\ref{eq.4.16}). We let $C \left(u_1, v_1, u_0\right)$ denote the gauge in general position. In the gauge $C \left(u_1 = \phi_{\eta}, v_1 = \phi_{\xi}, u_0 = \phi_{\xi \eta} + \phi_{\xi} \phi_{\eta}\right)$, which by definitions~(\ref{eq.4.4}) of the invariants corresponds to the zero values of the invariants $w_1$ and $w_2$,
\begin{equation}
w_1 = u_{1\xi} - v_{1\eta} = 0, \qquad w_2 = u_0 - u_{1\xi} - u_1 v_1 = 0, \qquad \widetilde w_2 = 0, \label{eq.4.17}
\end{equation}
system of $2D$gDLW equations~(\ref{eq.4.14})--(\ref{eq.4.16}) reduces to the two-dimensional B\"{u}rgers equation in potential form
\begin{equation}
\phi_t = -\kappa_1 \phi_{\xi \xi} - \kappa_2 \phi_{\eta \eta} - \kappa_1 (\phi_{\xi})^2 - \kappa_2 (\phi_{\eta})^2 + v_0, \label{eq.4.18}
\end{equation}
or, in terms of the variable $\rho$ related to the Hopf--Cole transformation $\phi = \log{\rho}$, to the linear diffusion equation
\begin{equation}
\rho_t = -\kappa_1 \rho_{\xi \xi} - \kappa_2 \rho_{\eta \eta} + v_0 \rho. \label{eq.4.19}
\end{equation}
It follows from the above construction that Eq.~(\ref{eq.4.18}) (or Eq.~(\ref{eq.4.19})) is the compatibility condition for auxiliary problems~(\ref{eq.4.1}) and (\ref{eq.4.2}) in Lax form,
\begin{equation}
[L_1, L_2] = B(w_1) L_1 \equiv 0. \label{eq.4.20}
\end{equation}

In another simple gauge, $C \left(u_1 = \phi_{\eta}, v_1 = 0, u_0 = 0\right)$, which by definitions~(\ref{eq.4.4}) corresponds to the invariants
\begin{equation}
w_1 = \phi_{\xi \eta}, \qquad w_2 = -\phi_{\xi \eta}, \qquad \widetilde w_2 = 0, \label{eq.4.21}
\end{equation}
system of $2D$gDLW equations~(\ref{eq.4.14})--(\ref{eq.4.16}) again reduces to the single B\"{u}rgers-type equation in potential form
\begin{equation}
\phi_t = \kappa_1 \phi_{\xi \xi} - \kappa_2 \phi_{\eta \eta} - \kappa_1 (\phi_{\xi})^2 + \kappa_2 (\phi_{\eta})^2 + v_0. \label{eq.4.22}
\end{equation}
This equation can be linearized by the Hopf--Cole transformation $\phi = -\log{\rho}$ to the corresponding linear equation
\begin{equation}
\rho_t = \kappa_1 \rho_{\xi \xi} - \kappa_2 \rho_{\eta \eta} - \rho v_0. \label{eq.4.23}
\end{equation}

In the gauge $C \left(u_1 = 0, v_1 = -q_{\xi} / q, u_0 = p \, q\right)$, it follows from~(\ref{eq.4.4}) that the invariants~$w_1$, $w_2$, and $\widetilde w_2$ are given by the expressions
\begin{equation}
w_1 = \bigl(\log{q}\bigr)_{\xi \eta}, \qquad w_2 = u_0 = p \, q, \qquad \widetilde w_2 = p \, q + \bigl(\log{q}\bigr)_{\xi \eta}, \label{eq.4.24}
\end{equation}
the variable $\rho$ according to~(\ref{eq.4.5}) has a constant value, and the variable $\phi$ is equal to zero. According to~(\ref{eq.4.14}), for the variable $v_0$, we have
\begin{equation}
v_0 = -2\kappa_1 \partial_{\eta}^{-1} w_{2\xi} = -2\kappa_1 \partial_{\eta}^{-1} \bigl(p \, q\bigr)_{\xi}. \label{eq.4.25}
\end{equation}
After some calculations in the case under study, system of $2D$gDLW equations (\ref{eq.4.14})--(\ref{eq.4.16}) implies the well-known system of DS equations~\cite{Davey&Stewartson_1974} for the variables $p$ and $q$:
\begin{equation}
\begin{split}
&p_t = \kappa_1 p_{\xi \xi} - \kappa_2 p_{\eta \eta} + 2\kappa_1 p \, \partial_{\eta}^{-1} \bigl(p \, q\bigr)_{\xi} - 2\kappa_2 p \, \partial_{\xi}^{-1} \bigl(p \, q\bigr)_{\eta}, \\
&q_t = -\kappa_1 q_{\xi \xi} + \kappa_2 q_{\eta \eta} - 2\kappa_1 q \, \partial_{\eta}^{-1} \bigl(p \, q\bigr)_{\xi} + 2\kappa_2 q \, \partial_{\xi}^{-1} \bigl(p \, q\bigr)_{\eta}.
\end{split} \label{eq.4.26}
\end{equation}

In the gauge $C \left(u_1 = p_{\eta}, v_1 = q_{\xi}, u_0 = p_{\eta} q_{\xi}\right)$, it follows from~(\ref{eq.4.4}) that the invariants can be expressed in terms of the variables $q$ and $p$ as
\begin{equation}
w_1 = p_{\xi \eta} - q_{\xi \eta}, \quad w_2 = -p_{\xi \eta}, \quad \widetilde w_2 = -q_{\xi \eta}. \label{eq.4.27}
\end{equation}
We substitute $w_1$, $w_2$, and $\widetilde w_2$ given by~(\ref{eq.4.27}) in system (\ref{eq.4.14})--(\ref{eq.4.16}) and obtain three equations for the variables $p$ and $q$. Equation~(\ref{eq.4.14}) for $\phi \equiv p$ implies the first equation
\begin{equation}
p_t = \kappa_1 p_{\xi \xi} - \kappa_2 p_{\eta \eta} - \kappa_1 (p_{\xi})^2 + \kappa_2 (p_{\eta})^2 - 2\kappa_2 p_{\eta} q_{\eta} + v_0. \label{eq.4.28}
\end{equation}
The other two equations follow from Eqs.~(\ref{eq.4.15}) and (\ref{eq.4.16}) for the invariants $w_2$ and $\widetilde w_2$ and can be expressed in terms of the variables $p$ and $q$ as
\begin{align}
&p_t = \kappa_1 p_{\xi \xi} - \kappa_2 p_{\eta \eta} - \kappa_1 (p_{\xi})^2 + \kappa_2 (p_{\eta})^2 + 2\kappa_1 \partial_{\eta}^{-1} \bigl(p_{\xi \eta} q_{\xi}\bigr) - 2\kappa_2 \partial_{\xi}^{-1} \bigl(p_{\xi \eta} q_{\eta}\bigr), \label{eq.4.29} \\
&q_t = -\kappa_1 q_{\xi \xi} + \kappa_2 q_{\eta \eta} + \kappa_1 (q_{\xi})^2 - \kappa_2 (q_{\eta})^2 - 2\kappa_1 \partial_{\eta}^{-1} \bigl(q_{\xi \eta} p_{\xi}\bigr) + 2\kappa_2 \partial_{\xi}^{-1} \bigl(q_{\xi \eta} p_{\eta}\bigr). \label{eq.4.30}
\end{align}
Equations~(\ref{eq.4.28}) and (\ref{eq.4.29}) are compatible under the choice of the variable
\begin{equation}
v_0 = 2\kappa_1 \partial_{\eta}^{-1} \bigl(p_{\xi \eta} q_{\xi}\bigr) + 2\kappa_2 \partial_{\xi}^{-1} \bigl(q_{\xi \eta} p_{\eta}\bigr). \label{eq.4.31}
\end{equation}
In this case, system of three equations (\ref{eq.4.28})--(\ref{eq.4.30}) reduces to system of two equations~(\ref{eq.4.29}) and (\ref{eq.4.30}), which contains the derivatives $p_{\xi \eta} q_{\xi}$, $p_{\xi \eta} q_{\eta}$, etc., in nonlocal terms.

Similarly, in the gauge $C \left(u_1 = p_{\eta}, v_1 = q_{\xi}, u_0 = 0\right)$, it follows from~(\ref{eq.4.4}) that the invariants $w_1$, $w_2$, and $\widetilde w_2$ can be expressed as
\begin{equation}
w_1 = p_{\xi \eta} - q_{\xi \eta}, \qquad w_2 = -p_{\xi \eta} - p_{\eta} q_{\xi}, \qquad \widetilde w_2 = -q_{\xi \eta} - p_{\eta} q_{\xi}. \label{eq.4.32}
\end{equation}
Equation~(\ref{eq.4.11}) for $\phi \equiv p$ with~(\ref{eq.4.32}) taken into account becomes
\begin{equation}
p_t = \kappa_1 p_{\xi \xi} - \kappa_2 p_{\eta \eta} - \kappa_1 (p_{\xi})^2 + \kappa_2 (p_{\eta})^2 - 2\kappa_2 p_{\eta} q_{\eta} + 2\kappa_1 \partial_{\eta}^{-1} \bigl(p_{\eta} q_{\xi}\bigr)_{\xi} + v_0. \label{eq.4.33}
\end{equation}
Substituting expressions~(\ref{eq.4.32}) for $w_1$ and $w_2$, we transform Eq.~(\ref{eq.4.12}) as
\begin{align}
&p_t - q_t = \kappa_1 \bigl(p + q\bigr)_{\xi \xi} - \kappa_2 \bigl(p + q\bigr)_{\eta \eta} - \kappa_1 \bigl(p_{\xi} - q_{\xi}\bigr)^2 + {} \nonumber \\
&\qquad {} + \kappa_2 \bigl(p_{\eta} - q_{\eta}\bigr)^2 + 2\kappa_1 \partial_{\eta}^{-1} \bigl(p_{\eta} q_{\xi}\bigr)_{\xi} - 2\kappa_2 \partial_{\xi}^{-1} \bigl(p_{\eta} q_{\xi}\bigr)_{\eta}. \label{eq.4.34}
\end{align}
Subtracting Eq.~(\ref{eq.4.34}) from (\ref{eq.4.33}), we obtain the evolution equation for the variable $q$:
\begin{equation}
q_t = -\kappa_1 q_{\xi \xi} + \kappa_2 q_{\eta \eta} + \kappa_1 \bigl(q_{\xi}\bigr)^2 - \kappa_2 \bigl(q_{\eta}\bigr)^2 - 2\kappa_1 p_{\xi} q_{\xi} + 2\kappa_2 \partial_{\xi}^{-1} \bigl(p_{\eta} q_{\xi}\bigr)_{\eta} + v_0. \label{eq.4.35}
\end{equation}
It follows from~(\ref{eq.4.32}) that Eq.~(\ref{eq.4.13}) for the invariant $w_2$ in terms of the variables $p$ and $q$ becomes
\begin{align}
&\bigl(p_{\xi \eta} + p_{\eta} q_{\xi}\bigr)_t = \kappa_1 \bigl(p_{\xi \eta} + p_{\eta} q_{\xi}\bigr)_{\xi \xi} - \kappa_2 \bigl(p_{\xi \eta} + p_{\eta} q_{\xi}\bigr)_{\eta \eta} - {} \nonumber \\
&\qquad {} - 2\kappa_1 \bigl((p_{\xi \eta} + p_{\eta} q_{\xi}) (p_{\xi} - q_{\xi})\bigr)_{\xi} + 2\kappa_2 \bigl((p_{\xi \eta} + p_{\eta} q_{\xi}) (p_{\eta} - q_{\eta})\bigr)_{\eta}. \label{eq.4.36}
\end{align}
Equations~(\ref{eq.4.34})--(\ref{eq.4.36}) are compatible if the variable $v_0$ satisfies the relation
\begin{equation}
v_{0\xi \eta} + p_{\eta} v_{0\xi} + q_{\xi} v_{0\eta} = 0. \label{eq.4.37}
\end{equation}
In the case where $v_0 \equiv 0$, system of three equations~(\ref{eq.4.34})--(\ref{eq.4.36}) implies the equivalent system of two equations
\begin{equation}
\begin{split}
&p_t = \kappa_1 p_{\xi \xi} - \kappa_2 p_{\eta \eta} - \kappa_1 \bigl(p_{\xi}\bigr)^2 + \kappa_2 \bigl(p_{\eta}\bigr)^2 - 2\kappa_2p_{\eta} q_{\eta} + 2\kappa_1 \partial_{\eta}^{-1} \bigl(p_{\eta} q_{\xi}\bigr)_{\xi}, \\
&q_t = -\kappa_1 q_{\xi \xi} + \kappa_2 q_{\eta \eta} + \kappa_1 \bigl(q_{\xi}\bigr)^2 - \kappa_2 \bigl(q_{\eta}\bigr)^2 - 2\kappa_1 p_{\xi} q_{\xi} + 2\kappa_2 \partial_{\xi}^{-1} \bigl(p_{\eta} q_{\xi}\bigr)_{\eta}.
\end{split} \label{eq.4.38}
\end{equation}
This system of equations was first obtained in a different context in~\cite{Konopelchenko_1988}.

To conclude this section, we discuss Miura-type transformations between different systems of nonlinear second-order DS-type equations obtained above in different gauges. For convenience, we let $P \equiv p$ and $Q \equiv q$ denote the solutions of system of nonlinear DS-type equations~(\ref{eq.4.26}). We use the invariants $w_1$ and $w_2$ to obtain the relations between the variables $P$ and $Q$ of DS system~(\ref{eq.4.26}) and the variables $p$ and $q$ of system~(\ref{eq.4.29}), (\ref{eq.4.30}):
\begin{equation}
w_1 = \bigl(\log{Q}\bigr)_{\xi \eta} = p_{\xi \eta} - q_{\xi \eta}, \qquad w_2 = P Q = -p_{\xi \eta}. \label{eq.4.39}
\end{equation}
Relation~(\ref{eq.4.39}) implies the corresponding Miura-type transformation
\begin{equation}
Q = e^{p - q}, \qquad P = -p_{\xi \eta} \, e^{q - p}. \label{eq.4.40}
\end{equation}

Quite similarly, for the pair of systems of DS-type equations~(\ref{eq.4.26}) and (\ref{eq.4.38}), we obtain
\begin{equation}
w_1 = \bigl(\log{Q}\bigr)_{\xi \eta} = p_{\xi \eta} - q_{\xi \eta}, \qquad w_2 = P Q = -p_{\xi \eta} - p_{\eta} q_{\xi}. \label{eq.4.41}
\end{equation}
From~(\ref{eq.4.41}), we also obtain a Miura-type transformation between the corresponding solutions
\begin{equation}
Q = e^{p - q}, \qquad P = -\bigl(p_{\xi \eta} + p_{\eta} q_{\xi}\bigr) e^{q - p}. \label{eq.4.42}
\end{equation}

Transformations~(\ref{eq.4.40}) and (\ref{eq.4.42}) permit obtaining solutions of the well-known system of DS equations~(\ref{eq.4.26}) from the corresponding solutions of systems of Eqs.~(\ref{eq.4.29}), (\ref{eq.4.30}), and (\ref{eq.4.38}). Therefore, these transformations are Miura-type transformations between the solutions of gauge-equivalent systems of second-order DS-type equations.

\section{Manifestly gauge-invariant system of integrable Nizhnik--Veselov--Novikov equations}
\setcounter{equation}{0}

We consider the results recently obtained in~\cite{Dubrovsky&Gramolin} concerning the gauge-invariant formulation of two-dimensional nonlinear evolution equations integrable using the two auxiliary linear problems
\begin{align}
&L_1 \psi = \bigl(\partial_{\xi \eta}^2 + u_1 \partial_{\xi} + v_1 \partial_{\eta} + u_0\bigr) \psi = 0, \label{eq.5.1} \\
&L_2 \psi = \bigl(\partial_t + \kappa_1\partial_{\xi}^3 + \kappa_2 \partial_{\eta}^3 + u_2 \partial_{\xi}^2 + v_2 \partial_{\eta}^2 + \tilde u_1 \partial_{\xi} + \tilde v_1 \partial_{\eta} + v_0\bigr) \psi = 0, \label{eq.5.2}
\end{align}
where $\kappa_1$ and $\kappa_2$ are constants, $\xi = x + \sigma y$ and $\eta = x - \sigma y$ are spatial coordinates, and $\sigma^2 = \pm 1$.

We use the compatibility condition in the Manakov triad form~\cite{Manakov_1976} (see~(\ref{eq.4.3})) to reduce auxiliary linear problems~(\ref{eq.5.1}) and (\ref{eq.5.2}) to the previously obtained and well-known Nizhnik--Veselov--Novikov (NVN) equations and to some other equations~\cite{Nizhnik_1980, Veselov&Novikov_1984, Konopelchenko_1990, Bogdanov_1987} (also see~\cite{Konopelchenko_book_2, Konopelchenko_book_3} and the references therein).

In the case under study, just as for the system of $2D$gDLW equations, formulas~(\ref{eq.4.4}) and (\ref{eq.4.5}) give the field variables, the classical invariants $w_2$ and $w_1$, and the pure gauge variable $\rho$, which are convenient for the gauge-invariant formulation. In the case of auxiliary linear problem~(\ref{eq.5.2}), the first invariant $w_1$ is equal to zero ($w_1 \equiv 0$), and the corresponding integrable system of nonlinear equations~\cite{Dubrovsky&Gramolin} in terms of the variables $\rho$ and $w_2$ has the form
\begin{align}
&\rho_t = -\kappa_1 \rho_{\xi \xi \xi} - \kappa_2 \rho_{\eta \eta \eta} - 3\kappa_1 \rho_{\xi} \partial_{\eta}^{-1} w_{2\xi} - 3\kappa_2 \rho_{\eta} \partial_{\xi}^{-1} w_{2\eta} + v_0 \rho, \label{eq.5.3} \\
&w_{2t} = -\kappa_1 w_{2\xi \xi \xi} - \kappa_2 w_{2\eta \eta \eta} - 3\kappa_1 \bigl(w_2 \partial_{\eta}^{-1} w_{2\xi}\bigr)_{\xi} - 3\kappa_2 \bigl(w_2 \partial_{\xi}^{-1} w_{2\eta}\bigr)_{\eta}. \label{eq.5.4}
\end{align}
System of equations~(\ref{eq.5.3}), (\ref{eq.5.4}) in terms of the variables $\phi = \log{\rho}$ and $w_2$ becomes
\begin{align}
&\phi_t = -\kappa_1 \phi_{\xi \xi \xi} - \kappa_2 \phi_{\eta \eta \eta} - \kappa_1 (\phi_{\xi})^3 - \kappa_2 (\phi_{\eta})^3 - 3\kappa_1 \phi_{\xi} \phi_{\xi \xi} - 3\kappa_2 \phi_{\eta} \phi_{\eta \eta} - {} \nonumber \\
&\qquad {} - 3\kappa_1 \phi_{\xi} \partial_{\eta}^{-1} w_{2\xi}- 3\kappa_2 \phi_{\eta} \partial_{\xi}^{-1} w_{2\eta} + v_0, \label{eq.5.5} \\
&w_{2t} = -\kappa_1 w_{2\xi \xi \xi} - \kappa_2 w_{2\eta \eta \eta} - 3\kappa_1 \bigl(w_2 \partial_{\eta}^{-1} w_{2\xi}\bigr)_{\xi} - 3\kappa_2 \bigl(w_2 \partial_{\xi}^{-1} w_{2\eta}\bigr)_{\eta}. \label{eq.5.6}
\end{align}

Equation~(\ref{eq.5.4}) (or (\ref{eq.5.6})) for the gauge invariant $w_2 = u_0 - u_{1\xi} - u_1 v_1$ of the last two systems exactly coincides in form with the well-known NVN equation~\cite{Nizhnik_1980, Veselov&Novikov_1984}
\begin{equation}
u_{t} = -\kappa_1 u_{\xi \xi \xi} - \kappa_2 u_{\eta \eta \eta} - 3\kappa_1 \bigl(u \partial_{\eta}^{-1} u_{\xi}\bigr)_{\xi} - 3\kappa_2 \bigl(u \partial_{\xi}^{-1} u_{\eta}\bigr)_{\eta}. \label{eq.5.7}
\end{equation}
Integrable system~(\ref{eq.5.3}), (\ref{eq.5.4}) (or (\ref{eq.5.5}), (\ref{eq.5.6})) can therefore be called the system of NVN equations.

Integrable system of NVN equations~(\ref{eq.5.3}), (\ref{eq.5.4}) (or (\ref{eq.5.5}), (\ref{eq.5.6})) has the following gauge structure:
\begin{itemize}
\item[\emph{a}.] It contains manifestly gauge-invariant equation~(\ref{eq.5.4}) (or (\ref{eq.5.6})) for the invariant~$w_2$.
\item[\emph{b}.] It contains Eq.~(\ref{eq.5.3}) (or (\ref{eq.5.5})) for the pure gauge variable $\rho$ (or $\phi$) with some additional terms containing the gauge invariant $w_2$ and the additional field variable $v_0$ used in auxiliary linear problem~(\ref{eq.5.2}).
\end{itemize}

For $w_2 = 0$, system of NVN equations~(\ref{eq.5.3}), (\ref{eq.5.4}) (or~(\ref{eq.5.5}), (\ref{eq.5.6})) reduces to the single linear equation
\begin{equation}
\rho_t = -\kappa_1 \rho_{\xi \xi \xi} - \kappa_2 \rho_{\eta \eta \eta} + v_0 \rho. \label{eq.5.8}
\end{equation}
In terms of the variable~$\phi = \log{\rho}$, linear equation~(\ref{eq.5.8}) looks like the third-order B\"{u}rgers equation
\begin{equation}
\phi_t = -\kappa_1 \phi_{\xi \xi \xi} - \kappa_2 \phi_{\eta \eta \eta} - \kappa_1 (\phi_{\xi})^3 - \kappa_2 (\phi_{\eta})^3 - 3\kappa_1 \phi_{\xi} \phi_{\xi \xi} - 3\kappa_2 \phi_{\eta} \phi_{\eta \eta} + v_0, \label{eq.5.9}
\end{equation}
which is linearized by the substitution $\phi = \log{\rho}$ and is hence $C$-integrable.

We let $C \left(\phi, u_0, v_0\right)$ denote the gauge corresponding to the nonzero field variables $u_1 = \phi_{\eta}$, $v_1 = \phi_{\xi}$, $u_0$, and $v_0$ of auxiliary linear problems~(\ref{eq.5.1}) and (\ref{eq.5.2}) and hence to system of NVN equations~(\ref{eq.5.5}), (\ref{eq.5.6}) in general position. In different gauges, the NVN system implies different gauge-equivalent integrable nonlinear equations. The solutions of these equations are related by Miura-type transformations.

For example, in the gauge $C \left(0, u_0, 0\right)$, system of NVN equations~(\ref{eq.5.5}), (\ref{eq.5.6}) reduces to the well-known NVN equation~\cite{Nizhnik_1980, Veselov&Novikov_1984} for the field variable $u_0$:
\begin{equation}
u_{0t} = -\kappa_1 u_{0\xi \xi \xi} - \kappa_2 u_{0\eta \eta \eta} - 3\kappa_1 \bigl(u_0 \partial_{\eta}^{-1} u_{0\xi}\bigr)_{\xi} - 3\kappa_2 \bigl(u_0 \partial_{\xi}^{-1} u_{0\eta}\bigr)_{\eta}. \label{eq.5.10}
\end{equation}

In another gauge, $C \left(\phi, 0, v_0\right)$, NVN system~(\ref{eq.5.5}), (\ref{eq.5.6}) becomes
\begin{align}
&\phi_t = -\kappa_1 \phi_{\xi \xi \xi} - \kappa_2 \phi_{\eta \eta \eta} - \kappa_1 (\phi_{\xi})^3 - \kappa_2 (\phi_{\eta})^3 + \nonumber \\
&\qquad {} + 3\kappa_1 \phi_{\xi} \partial_{\eta}^{-1} \bigl(\phi_{\xi} \phi_{\eta}\bigr)_{\xi} + 3\kappa_2 \phi_{\eta} \partial_{\xi}^{-1} \bigl(\phi_{\xi} \phi_{\eta}\bigr)_{\eta} + v_0, \label{eq.5.11} \\
&\bigl(\partial_{\xi \eta}^2 + \phi_{\eta} \partial_{\xi} + \phi_{\xi} \partial_{\eta}\bigr) \phi_t = \bigl(\partial_{\xi \eta}^2 + \phi_{\eta} \partial_{\xi} + \phi_{\xi} \partial_{\eta}\bigr) \Bigl[-\kappa_1 \phi_{\xi \xi \xi} - \kappa_2 \phi_{\eta \eta \eta} - \nonumber \\
&\qquad {} - \kappa_1 (\phi_{\xi})^3 - \kappa_2 (\phi_{\eta})^3 + 3\kappa_1 \phi_{\xi} \partial_{\eta}^{-1} \bigl(\phi_{\xi} \phi_{\eta}\bigr)_{\xi} + 3\kappa_2 \phi_{\eta} \partial_{\xi}^{-1} \bigl(\phi_{\xi} \phi_{\eta}\bigr)_{\eta}\Bigr]. \label{eq.5.12}
\end{align}
By~(\ref{eq.5.11}) and (\ref{eq.5.12}), NVN system~(\ref{eq.5.5}), (\ref{eq.5.6}) in the gauge $C \left(\phi, 0, v_0\right)$ reduces to the system of equations
\begin{equation}
\begin{split}
&\phi_t = -\kappa_1 \phi_{\xi \xi \xi} - \kappa_2 \phi_{\eta \eta \eta} - \kappa_1 (\phi_{\xi})^3 - \kappa_2 (\phi_{\eta})^3 + {} \\
&\qquad {} + 3\kappa_1 \phi_{\xi} \partial_{\eta}^{-1} \bigl(\phi_{\xi} \phi_{\eta}\bigr)_{\xi} + 3\kappa_2 \phi_{\eta} \partial_{\xi}^{-1} \bigl(\phi_{\xi} \phi_{\eta}\bigr)_{\eta} + v_0, \\
&\bigl(\partial_{\xi \eta}^2 + \phi_{\eta} \partial_{\xi} + \phi_{\xi} \partial_{\eta}\bigr) v_0 = 0.
\end{split} \label{eq.5.13}
\end{equation}
For $v_0 = 0$, system of equations~(\ref{eq.5.13}) reduces to the well-known modified NVN (mNVN) equation
\begin{equation}
\phi_t = -\kappa_1 \phi_{\xi \xi \xi} - \kappa_2 \phi_{\eta \eta \eta} - \kappa_1 (\phi_{\xi})^3 - \kappa_2 (\phi_{\eta})^3 + 3\kappa_1 \phi_{\xi} \partial_{\eta}^{-1} \bigl(\phi_{\xi} \phi_{\eta}\bigr)_{\xi} + 3\kappa_2 \phi_{\eta} \partial_{\xi}^{-1} \bigl(\phi_{\xi} \phi_{\eta}\bigr)_{\eta}, \label{eq.5.14}
\end{equation}
which was first obtained in a different context by Konopelchenko~\cite{Konopelchenko_1990}. We note that mNVN equation~(\ref{eq.5.14}) differs from the mNVN equation obtained by Bogdanov~\cite{Bogdanov_1987}. The new system of equations~(\ref{eq.5.13}) can also be called a system of mNVN equations.

Obviously, the solutions $u_0$ and $\phi$ of NVN equations~(\ref{eq.5.10}) and mNVN equations~(\ref{eq.5.14}) are related by the Miura-type transformation
\begin{equation}
u_0 = -\phi_{\xi \eta} - \phi_{\xi} \phi_{\eta} \label{eq.5.15}
\end{equation}
through the gauge invariant $w_2 = u_0 = -\phi_{\xi \eta} - \phi_{\xi} \phi_{\eta}$ (calculated in the different gauges $C \left(0, u_0, 0\right)$ and $C \left(\phi, 0, 0\right)$).

In the one-dimensional limit with coinciding derivatives $\partial_{\xi} = \partial_{\eta}$, mNVN equation~(\ref{eq.5.14}) reduces to mKdV equation in potential form
\begin{equation}
\phi_t = -\kappa \, \phi_{\xi \xi \xi} + 2\kappa (\phi_{\xi})^3, \label{eq.5.16}
\end{equation}
where $\kappa = \kappa_1 + \kappa_2$. In terms of the variable $v_1 = \phi_{\xi}$, this is the mKdV equation
\begin{equation}
v_{1t} = -\kappa \, v_{1\xi \xi \xi} + 6\kappa \, v_1^2 v_{1\xi}. \label{eq.5.17}
\end{equation}

\section{Conclusion}
\setcounter{equation}{0}

The ideas of gauge invariance are currently widely used in the theory of integrable nonlinear equations. A manifestly gauge-invariant formulation of integrable systems of nonlinear equations can be given in all cases where the gauge freedom is properly taken into account in the corresponding auxiliary linear problems.

\paragraph{Acknowledgments.}

This work is supported by the Analytic Departmental Goal-Oriented Program of the Russian Federation Ministry of Science and Education ``Development of the Higher School Potential (2009--2010)'' (Grant No. 2.1.1/1958) and Novosibirsk State Technical University (Grant in Basic Research 2008--2009).

\end{document}